\documentclass[journal=jacsat,manuscript=article]{achemso}
\usepackage[version=3]{mhchem}
\setkeys{acs}{doi=true}
\setkeys{acs}{maxauthors=100}

\usepackage[english]{babel}
\usepackage{hyperref}
\hypersetup{
    pdfstartview={FitH},
    pdfpagemode={UseNone},
    colorlinks=true,  % Keep hyperlinks clickable but set all colors to black
    linkcolor=black,  % Internal document links (e.g., figures, sections)
    citecolor=black,  % Citations
    urlcolor=black,   % URLs
    bookmarksopen=true,
    pdfnewwindow=true
}
\usepackage[all]{hypcap}  % Ensures links point to the top of figures/tables

\usepackage{chemformula} % Formula subscripts using \ch{}
\usepackage[T1]{fontenc} % Use modern font encodings
\usepackage{braket}

\title[Enhanced photon-pair generation from a van der
Waals metasurface]
  {Enhanced photon-pair generation from a van der
Waals metasurface}

\author{Tongmiao Fan}
\affiliation
{ARC Centre of Excellence for Transformative Meta-Optical Systems (TMOS), Department of Electronic Materials Engineering, Research School of Physics, The~Australian National University, Canberra, ACT 2601, Australia}
\altaffiliation{Co-first authors with equal contribution}

\author{Yilin Tang}
\affiliation{School of Engineering, College of Engineering, Computing and Cybernetics,The~Australian National University, Canberra, ACT 2601, Australia}
\alsoaffiliation{ARC Centre for Quantum Computation and Communication Technology, The Australian National University, Canberra, ACT 2601, Australia}
\altaffiliation{Co-first authors with equal contribution}

\author{Shaun Lung}
\affiliation{Institute of Applied Physics, Abbe Center of Photonics, Friedrich Schiller University Jena, Jena 07745, Germany}

\author{Maximilian Weissflog}
\affiliation
{Institute of Applied Physics, Abbe Center of Photonics, Friedrich Schiller University Jena, Jena 07745, Germany}

\author{Jinyong Ma}
\affiliation
{ARC Centre of Excellence for Transformative Meta-Optical Systems (TMOS), Department of Electronic Materials Engineering, Research School of Physics, The~Australian National University, Canberra, ACT 2601, Australia}
\alsoaffiliation{Institute of Quantum Precision Measurement, State Key Laboratory of Radio Frequency Heterogeneous Integration, College of Physics and Optoelectronic Engineering, Shenzhen University, Shenzhen 518060, P. R. China}

\author{Saniya Shinde}
\affiliation
{Institute of Applied Physics, Abbe Center of Photonics, Friedrich Schiller University Jena, Jena 07745, Germany}

\author{Sina Saravi}
\affiliation
{Institute of Applied Physics, Abbe Center of Photonics, Friedrich Schiller University Jena, Jena 07745, Germany}

\author{Mudassar Nauman}
\affiliation
{BluGlass Ltd, Sydney, NSW 2128, Australia}

\author{Wenkai Yang}
\affiliation
{School of Engineering, College of Engineering, Computing and Cybernetics,
The Australian National University, Canberra, ACT 2601, Australia}

\author{Hao Qin}
\affiliation
{School of Engineering, College of Engineering, Computing and Cybernetics,
The Australian National University, Canberra, ACT 2601, Australia}

\author{Shuyao Qiu}
\affiliation
{School of Engineering, College of Engineering, Computing and Cybernetics,
The Australian National University, Canberra, ACT 2601, Australia}

\author{Andrey A. Sukhorukov}
\email{andrey.sukhorukov@anu.edu.au}
\affiliation
{ARC Centre of Excellence for Transformative Meta-Optical Systems (TMOS), Department of Electronic Materials Engineering, Research School of Physics, The~Australian National University, Canberra, ACT 2601, Australia}

\author{Yuerui Lu}
\email{yuerui.lu@anu.edu.au}
\affiliation
{School of Engineering, College of Engineering, Computing and Cybernetics,
The Australian National University, Canberra, ACT 2601, Australia}
\alsoaffiliation{ARC Centre for Quantum Computation and Communication Technology, The Australian National University, Canberra, ACT 2601, Australia}

\author{Frank Setzpfandt}
\email{f.setzpfandt@uni-jena.de}
\affiliation
{Institute of Applied Physics, Abbe Center of Photonics, Friedrich Schiller University Jena, Jena 07745, Germany}
\alsoaffiliation{Fraunhofer Institute for Applied Optics and Precision Engineering IOF, Jena 07745, Germany}

\begin{document}
% \sloppy
% \restoregeometry

%%%%%%%%%%%%%%%%%%%%%%%%%%%%%%%%%%%%%%%%%%%%%%%%%%%%%%%%%%%%%%%%%%%%%
%% The "tocentry" environment can be used to create an entry for the
%% graphical table of contents. It is given here as some journals
%% require that it is printed as part of the abstract page. It will
%% be automatically moved as appropriate.
%%%%%%%%%%%%%%%%%%%%%%%%%%%%%%%%%%%%%%%%%%%%%%%%%%%%%%%%%%%%%%%%%%%%%

%%%%%%%%%%%%%%%%%%%%%%%%%%%%%%%%%%%%%%%%%%%%%%%%%%%%%%%%%%%%%%%%%%%%%
%% The abstract environment will automatically gobble the contents
%% if an abstract is not used by the target journal.
%%%%%%%%%%%%%%%%%%%%%%%%%%%%%%%%%%%%%%%%%%%%%%%%%%%%%%%%%%%%%%%%%%%%%
\begin{abstract}
Quantum photon pairs play a pivotal role in many quantum applications. Metasurfaces, two-dimensional arrays of nanostructures, have been studied intensively to enhance and control pair generation via spontaneous parametric downconversion (SPDC). Van der Waals (VdW) layered materials have emerged as promising candidates for nonlinear materials in quantum light sources, owing to their high nonlinear susceptibility and compatibility with on-chip integration. In this work, we present the first demonstration of SPDC from a metasurface composed of the VdW material 3R-MoS$_2$. The nanoresonators support quasi-bound states in the continuum (qBIC) with a quality factor of up to 120, enhancing light-matter interactions. This design achieves a 20-fold increase in SPDC rate compared to an unstructured film and significantly higher brightness, resulting in enhanced quantum photon-pair generation. This work establishes a new approach for utilizing van der Waals metasurfaces in the generation of quantum photon pairs, opening avenues for advanced quantum applications.
 
%We also demonstrated ultimate enhancement of SPDC process based on multiple resonance. 
%It not only makes use of the geometry resonance from the nanostructure, but also exciton resonance from 3R-MoS$_2$, which provides one of the highest second order 
\end{abstract}

%%%%%%%%%%%%%%%%%%%%%%%%%%%%%%%%%%%%%%%%%%%%%%%%%%%%%%%%%%%%%%%%%%%%%
%% Start the main part of the manuscript here.
%%%%%%%%%%%%%%%%%%%%%%%%%%%%%%%%%%%%%%%%%%%%%%%%%%%%%%%%%%%%%%%%%%%%%
%\section{Introduction}
Quantum-correlated pairs of photons are of great interest due to their numerous applications in emerging quantum technologies such as quantum computing~\cite{Slussarenko:2019-41303:APPR}, communication~\cite{Li2004}, and imaging~\cite{Clark:2021-60401:APL}. 
A widely used approach to generate photon pairs is spontaneous parametric down-conversion (SPDC), which is enabled by the second-order nonlinear susceptibility $\chi^{(2)}$ of non-centrosymmetric materials.
SPDC spontaneously splits photons from a pump laser into correlated pairs of photons at lower frequencies, called signal (s) and idler (i) photons. 
Conventionally, bulk nonlinear crystals are used for SPDC~\cite{Klyshko:1988:PhotonsNonlinear, Couteau:2018-291:CTMP}. However, the need for phase matching to achieve efficient photon-pair generation in these crystals creates challenges for their application, as the dispersion of naturally occurring crystals limits the frequency ranges where phase matching for efficient SPDC can be achieved.
Alternative techniques like quasi-phase-matching are technologically challenging and restricted to specific materials. 
%However, their large size and the need for temperature-control elements complicate integration into compact, scalable optical devices~\cite{Wang:2020-273:NPHOT}.

To circumvent these challenges, significant attention has recently been devoted to developing ultrathin sources for photon pairs.
Nonlinear thin films relax the longitudinal phase-matching condition for SPDC to make larger spectral and angular ranges accessible for the generated pairs~\cite{Okoth:2020-11801:PRA,Santiago-Cruz:2021-653:OL,Sultanov:2022-3872:OL}. Thus, they enhance the flexibility of photon-pair generation, and due to their essentially two-dimensional geometry they can be easily integrated in many optical systems.

 %to enhance integration capabilities and introduce extra novel functionalities

First experiments on thin-film photon-pair sources focused on thin crystalline layers of materials also commonly employed for SPDC in bulk crystals~\cite{Okoth:2020-11801:PRA,Santiago-Cruz:2021-653:OL,Sultanov:2022-3872:OL}. The ultimate limit of the thickness can be reached using two-dimensional materials, of which in particular transition-metal dichalcogenides (TMDCs) feature a large second-order nonlinearity~\cite{Zograf:2024-751:NPHOT} (see Sec.~S6 in supplementary for detailed comparison). However, photon-pair generation in monolayer TMDC crystals with only one molecular layer could not unequivocally be demonstrated experimentally~\cite{Guo:2023-53:NAT,Xu:2022-698:NPHOT}. Van der Waals (VdW) crystals are stacks of two-dimensional crystals connected only by weak VdW forces. In such a VdW crystal made from niobium oxide dichloride (NbOCl$_2$) with a thickness of several hundred nanometers, the generation of photon pairs was demonstrated experimentally~\cite{Guo:2023-53:NAT}. 
VdW crystals consisting of TMDCs can preserve the second-order nonlinearity of the monolayer crystal only in the 3R-configuration, where the strength of the nonlinear interaction can be scaled by controlling the thickness~\cite{Autere2018AM, Tang2024NC}. 
Such 3R-TMDC crystals have been implemented as photon-pair sources, and, due to their specific crystalline symmetry, enable the direct generation of polarization-entangled pairs~\cite{Weissflog:2024-7600:NCOM, Feng:2024-16:ELI, Kallioniemi:2025-142:NPHOT, Lyu2024NC}.

%Various material systems have been explored as platforms for quantum photon pairs~\cite{Poddubny:2016-123901:PRL, Guo:2023-53:NAT}. 
%Among these, submicron thin films made up of van der Waals  (VdW) crystals are of great interest due to their numerous advantages.
%For example, VdW can easily integrate with existing technologies, generate tunable polarization entanglement, and allow scalable second-order optical nonlinearity~\cite{Autere2018AM, Weissflog:2024-7600:NCOM, Feng:2024-16:ELI, Tang2024NC, Kallioniemi:2025-142:NPHOT}. 
%These integration capabilities and scalability are crucial for the development of quantum technologies.

While nonlinear thin films offer significant advantages for the SPDC process, their practical applications are still limited, as the achieved photon-pair generation rates are rather low.
To address this issue, researchers have studied the resonant enhancement of SPDC using dielectric nanoresonators, providing field enhancement for the interacting modes~\cite{Marino2019, Nikolaeva:2021-43703:PRA, Duong:2022-3696:OME, Zilli2023, Saerens:2023-3245:NANL, Olekhno:2024-245416:PRB, Weissflog:2024-11403:APPR}. 
It was experimentally demonstrated that even single nanoresonators can generate photon pairs. 
However, due to the small volume and limited field enhancement achievable in individual nanostructures, the generation rates were still low.
%For example, it has been reported a nanoresonator made up of AlGaAs to maximize the nonlinear rate by combining Mie resonances at both the pump and bi-photon wavelengths~\cite{Marino2019}.
%However, a single nanoresonator lacks comprehensive control over light and requires a complex setup to focus the beam, limiting scalability and integration capabilities.

Metasurfaces, which are arrays of nanoresonators~\cite{Kuznetsov:2024-816:ACSP, Schulz:2024-260701:APL}, can enhance the nonlinear process while preserving scalability and integration capabilities comparable to thin films~\cite{Santiago-Cruz:2021-4423:NANL}. Besides enhancing the generation rate by simply combining many nanoresonators, metasurfaces also allow one to exploit collective resonant effects, like quasi-bound states in the continuum (qBIC) resonances~\cite{Koshelev2018PRL}, which offer much larger field enhancement and can further boost the conversion efficiency. Such structures possessing resonances with high quality factors have been demonstrated to enhance the photon-pair generation rate~\cite{Santiago-Cruz:2022-991:SCI} while enabling precise control of the generated wavelengths due to the narrow bandwidth of the resonances. Furthermore, control of the polarization~\cite{Ma:2023-8091:NANL,Jia:2025-eads3576:SCA} and spatial~\cite{Weissflog:2024-3563:NANP} degrees of freedom of the generated photon pairs was demonstrated using metasurfaces.

%More interestingly, nonlinear metasurfaces can modify light in different degrees of freedom to generate tailored quantum states that can be applied in a wide range of quantum applications.
%For example, there have been a few papers on nonlinear metasurfaces based on quasi-bound states in the continuum (qBIC) resonance by breaking the symmetry~\cite{Koshelev2018PRL}. 
%Such a resonance of a high quality factor can greatly enhance the electric field in the nanoresonators, leading to a boosted photon pair generation rate.
%Additionally, thanks to the narrow bandwidth of the resonance, it can be used to generate complex quantum states that cannot be easily achieved with classical counterparts~\cite{Santiago-Cruz:2022-991:SCI}.

To build on these results, new material systems for metasurfaces to further enhance SPDC are in urgent demand. 
Here, VdW materials, especially TMDCs, hold a lot of promise due to their high refractive index, which is advantageous for achieving field enhancement in nanoresonators. 
Their unique nonlinear tensor carrying a large nonlinear coefficient enables polarization engineering with higher SPDC efficiency.
Furthermore, their high damage threshold surpasses that of often used III-V semiconductors, facilitating the use of larger pump powers in SPDC~\cite{Xu:2022-698:NPHOT}.

\begin{figure*}[!b] \centering
    \includegraphics[width=0.98\textwidth]{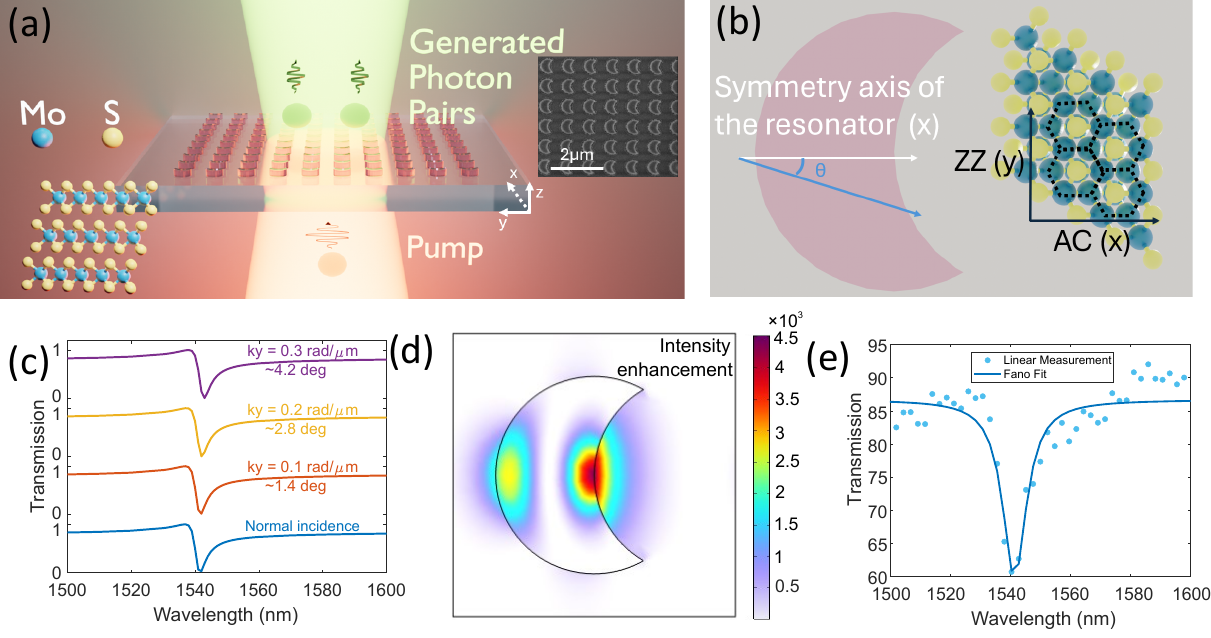} 
    \caption{
    \textbf{a},~Sketch of photon-pair generation from a 3R-MoS$_2$ metasurface. The left inset shows the molecular structure of 3R-MoS$_2$, the right inset is an SEM image of the fabricated metasurface.
    \textbf{b}, The orientation of the crystal structure and the nanoresonators. The symmetry axis of the nanoresonators overlaps with the ZZ axis of the 3R-MoS$_2$ crystal.
    \textbf{c},~The simulated transmission for different incident angles. The angular dispersion is small according to the shift of the resonant wavelength.
    \textbf{d},~Cross-section of the scattered electric field intensity enhancement in the middle of the resonator (xy-plane) at the resonant wavelength of 1540~nm.
    \textbf{e},~Experimentally measured transmission. A qBIC resonance with a quality factor of 120 is observed around 1540~nm.
}
\label{fig:1}
\end{figure*}

%To the best of our knowledge, the photon pair generation from metasurfaces made up of nonlinear VdW remains unexplored. 
Here, we experimentally realized photon-pair generation by SPDC from a VdW metasurface made from 3R-MoS$_2$ as schematically shown in Fig.~\ref{fig:1}(a). To leverage the advantageous properties of this material, we designed a metasurface supporting a qBIC resonance with a quality factor of 120 to enhance light-matter interaction. This increases the overall photon-pair count rate compared to a thin film from the same material by a factor of 20, corresponding to a larger enhancement in the spectral brightness across the narrow bandwidth of the resonance.
%The 3R-MoS$_2$ phase exhibits several advantageous properties that make it highly suitable for quantum photon pair generation from metasurfaces. 
%It exhibits a high refractive index, making it ideal for high-Q metasurfaces, and a large second-order susceptibility, as previously noted.  
%Furthermore, 3R-MoS$_2$ demonstrates an exceptionally high damage threshold ($>$45 GW/cm² under 1550 nm femtosecond laser pumping, based on our measurements), greatly surpassing that of III-V semiconductors.
%In our work, polarization-dependent SHG measurement was done to analyze the in-plane lattice-symmetry of 3R-MoS$_2$. 
%The specially designed metasurface supports a quasi-bound states in the continuum (qBIC) resonance with a quality factor of 150, which greatly enhances the light-matter interaction~\cite{Koshelev2018PRL, Santiago-Cruz:2022-991:SCI}. 
%Therefore, a 20-fold enhancement in the overall rate is observed with a higher coincidence to accidental (CAR) ratio compared to thin films. 
%Thanks to the narrow bandwidth, a much higher brightness enhancement is achieved. 
%Such great brightness enhancement allows the application of the metasurface in various technologies like quantum sensing.
%We also demonstrate that the generated quantum states can be tailored via the orientation of resonators to adapt the used elements of the nonlinear tensor to sepcific applications. % which can overcome the limitation of settled second-order nonlinear tensor.
Our work paves the way for the application of VdW crystals for the generation of quantum photon pairs from nonlinear metasurfaces.
%Furthermore, we also achieve ultimate enhancement of quantum photon pair generation rate via combining the geometry resonance with the material resonance.

%\subsection*{Concept and modeling}
We developed a resonant metasurface from a VdW material for the generation of photon pairs by SPDC. 
The resonators are created by etching a film of 3R-MoS$_2$ situated on a sapphire substrate, as shown in Fig.~\ref{fig:1}(a). To support a high-Q resonance that enhances the light-matter interaction, we designed and fabricated a metasurface with a broken in-plane inversion symmetry due to crescent-shaped nanoresonators in square lattices, as shown in the scanning electron microscope (SEM) image in Fig.~\ref{fig:1}(a) and in Fig.~\ref{fig:1}(b).
The detailed fabrication procedure is shown in Section S7 in supplementary.
The period of one unit cell is set to 790~nm, and the thickness of the resonators (MoS$_2$) is 222~nm.
The diameter of the crescent is set to 551~nm to support a qBIC resonance at 1540~nm (see detailed design procedure in  Sec.~S1, Supplementary).
Multipole decomposition presented in~\cite{nauman2025dynamic} confirms that the crescent-shaped geometry supports an ideal magnetic dipole mode, as the electric dipole (ED) and toroidal dipole (TD) contributions cancel each other at the same spectral position as the magnetic dipole mode. 
As shown in Fig.~\ref{fig:1}(b), the symmetry axis ($x$) of the crescent-shaped nanoresonators is aligned with the armchair (AC) direction of the MoS$_2$ crystal, while the zigzag (ZZ) direction of the crystal aligns with the $y$-axis.
%With specially oriented nanoresonators, we are able to tailor the generated polarization states by engineering the electric field at resonance.
%The right inset in Fig. \ref{fig:1}(a) shows a scanning electron microscope (SEM) image of our sample.
%The period of one unit cell is set to 790~nm; and the thickness resonators is 222~nm.
Due to the specific crystal orientation and non-centrosymmetric nature of 3R-MoS$_2$ as indicated in Fig. \ref{fig:1}(a) and (b), it has $\chi^{(2)}$ components $\chi_{y y y}^{(2)}=-\chi_{y x x}^{(2)}=-\chi_{x x y}^{(2)}=-\chi_{x y x}^{(2)}$ in the nonlinear tensor. Here, we target the generation of co-polarized photon pairs, where the polarizations of signal, idler, and pump are aligned along the y-direction.
Such a combination of polarizations enables the maximal SPDC rate as analyzed in Supplementary (Sec.~S1.3).
%Those components enable the tuning of quantum polarization-entangled states by varying pump polarization from thin films~\cite{Weissflog:2024-7600:NCOM}.
%However, the generated quantum states are always restricted by the nonlinear tensor of 3R-MoS$_2$.
%From a metasurface made from arrays of nanoresonators, we have better flexibility to tailor the generated polarization states~\cite{Ma:2025-adu4133:SCA}. 
%The right inset in Fig. \ref{fig:1}(a) shows a scanning electron microscope (SEM) image of our sample.

As indicated in Fig.~\ref{fig:1}(c), the simulated transmission spectra at different incident angles indicate that the resonant wavelengths are almost independent of the incident angles within a range of several degrees, in contrast to nonlocal modes~\cite{Liang:2024-53801:PRL, sci_jihua}.
It is also observed that the quality factors remain at a high value even at an angle of 0.2~rad/$\mu$m ($\approx$3 degrees for a photon of 1540~nm).
This indicates a broad emission angle of photon pair generation at resonance due to the excitation of the localized mode.
The broadly emitted SPDC is favorable in many applications, like quantum imaging~\cite{Ma:2025-2:ELI}.

The transmission dips around 1540~nm are caused by a resonance. Fig.~\ref{fig:1}(d) shows the enhancement of the dominating resonant intensity component of the qBIC mode, $|E_y|^2$,  relative to the incident plane wave.
The high refractive index of 3R-MoS$_2$ ($\approx$ 4.6 at 1540~nm) enhances the ability to trap light compared to other material systems such as III-V semiconductors~\cite{Zograf:2024-751:NPHOT, Xu:2022-698:NPHOT}.
Combined with the specially designed nanoresonators that support a sharp qBIC resonance, the metasurface enables an intensity enhancement of up to 4.5$\times10^3$ times for y-polarized field components inside the nonlinear 3R-MoS$_2$.
Such a high local intensity enhancement can greatly improve the probability of photon-pair generation in the resonant spectral band and hence, the overall SPDC rate. 
Introducing resonances for certain polarization directions also influences the generated polarization quantum state. For a non-resonant system like a thin film, the symmetry of the 3R-MoS$_2$ nonlinear tensor allows one to directly generate entangled polarization states.
Taking, for instance, a y-polarized pump photon, down-conversion via the two tensor elements $\chi^{(2)}_{yyy}=-\chi^{(2)}_{yxx}$ generates a maximally entangled Bell state $\frac{1}{\sqrt{2}}(\ket{HH}-\ket{VV})$. 
In a resonant system like the qBIC metasurface, the situation is different: the increased density of states at the y-polarized high-Q resonance strongly increases the down-conversion probability into a pair of y-polarized photons compared to a pair of x-polarized photons, leading to a non-entangled $\ket{VV}$ state.
%Meanwhile, the enhancement for a specific E component can transfer the generated state from a polarization-entangled Bell state to an unentangled co-polarized state due to its effect on the nonlinear polarization component $P_y=\epsilon_0(E_{x,s}E_{x,i}\chi_{y x x}^{(2)}+E_{y,s}E_{y,i}\chi_{y y y}^{(2)}+E_{y,s}E_{z,i}\chi_{y y z}^{(2)}+E_{z,s}E_{y,i}\chi_{y z y}^{(2)})$.

To confirm the simulated linear optical properties, we experimentally measured the transmission through the metasurface at a normal incidence angle, which is plotted in Fig.~\ref{fig:1}(e).
As we can see, the fabricated metasurface supports a sharp resonance around 1540~nm, whose quality factor is around 120 based on Fano fitting of the transmission spectrum.
The corresponding linewidth is around 10~nm.
In a bulk 3R-MoS$_2$ crystal of the same thickness, no resonant feature is observed (see Sec.~S2 in Supplementary).
The measured transmission verifies the effectiveness of our design and the quality of the fabrication process.
%Interestingly, the metasurface also supports a low-Q resonance around two times of the exciton wavelength ($\approx$1249~nm).
%It reveals the potential to maximize the photon pair generation rate via combining the qBIC resonance with the material resonance, which has never been explored before.

%\subsection*{Enhanced Second Harmonic Generation (SHG) in 3R-MoS$_2$ Metasurfaces}

\begin{figure*}[!b] \centering
    \includegraphics[width=1\textwidth]{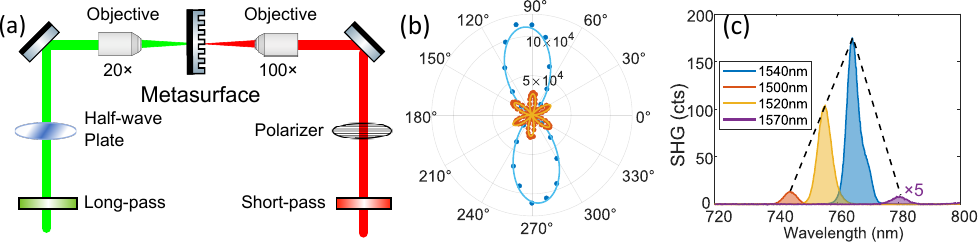} 
    \caption{
    \textbf{a}, Schematic diagram of the experimental setup for SHG measurement of 3R-MoS$_2$ qBIC metasurface. In the experiment, the polarization of the incident laser was initially set to be along the armchair direction of the flake and controlled by rotating the $\lambda$/2 waveplate. A linear polarizer was used to select the polarization component of the SH radiation parallel to the polarization of the pump beam. 
    \textbf{b}, Measured (dots) co-polarized SHG counts and fitting (solid line) of the counts as a function of the polarization angle of the incident laser, from 3R-MoS$_2$ metasurface (blue color), and 3R-MoS$_2$ flake (orange). 
    \textbf{c}, Measured SHG spectra for different input pulse central wavelengths from the metasurface. The maximal points are connected through a dashed line.
}
\label{fig:2}
\end{figure*}

We present in Fig.~\ref{fig:2} the results of the comprehensive characterization of second-harmonic generation (SHG) in the realized crescent-structured qBIC 3R-MoS$_2$ metasurface. The scheme in Fig.~\ref{fig:2}(a) illustrates the experimental configuration for SHG measurements, with details provided in the methods section in supplementary (S8). 
A polarization dependent SHG measurement was performed on the 3R-MoS$_2$ unpatterned film, demonstrating the lattice orientation and serving as a reference for comparison. The SHG intensity in the polarization direction parallel to the excitation polarization exhibits a characteristic six-fold symmetry, consistent with the hexagonal crystal symmetry of 3R-MoS$_2$ (orange dots in Fig.~\ref{fig:2}(b)). The maxima correspond to excitation and generation along the armchair direction of MoS$_2$. The SHG intensity from the metasurface (blue dots in Fig.~\ref{fig:2}(b)) exhibits a qualitatively different polarization dependence when measured in the same configuration. Here we found only two maxima, which correspond to the polarization direction exciting the qBIC resonance. The signal was maximized along the polarization y-axis ($\theta$ = 90$^{\circ}$) of the pump incident, aligning with the dominant qBIC resonance mode, which was also revealed in the transmission curve (Fig.~\ref{fig:1}(e)).
The definition of $\theta$ can be found in Fig. \ref{fig:1}(b).
A slight shift is observed between the maxima of the unpatterned film and the metasurface. 
This shift can be attributed to imperfections in the fabrication of the resonator, which caused misalignment between its symmetry axis and the crystal's AC axis.
The SHG emission intensity along the polarization direction of maximum conversion efficiency is significantly higher than for the unstructured 3R-MoS$_2$, thus confirming the ability of the metasurface to enhance nonlinear interaction.
%It further confirms the signature that differs from the six-fold symmetry observed in the bulk material, demonstrating the capacity of the metasurface to manipulate the nonlinear response.

To demonstrate the resonant nature of the SHG enhancement, we measured spectrally resolved SH intensities for different central wavelengths of the fundamental beam. The measurement results are shown in Fig.~\ref{fig:2}(c). A significant increase in SHG intensity is observed as the excitation wavelength approaches the qBIC resonance at 1540 nm with $\theta$ = 90$^{\circ}$. At the resonant wavelength, the metasurface exhibits an approximately 90-fold increase in SHG intensity compared to the off-resonance position at 1570 nm. The observed enhancement in SHG intensity can be attributed to the local field enhancement enabled by the qBIC resonance~\cite{Koshelev2018PRL}. 
The symmetry-broken metasurface effectively couples the incident light to the qBIC mode and contributes to the light confinement, leading to a significant enhancement of the nonlinear response.

%\subsection*{Experimental characterization of enhanced photon-pair generation from qBIC resonance}

\begin{figure*}[!b] \centering
    \includegraphics[width=0.8\textwidth]{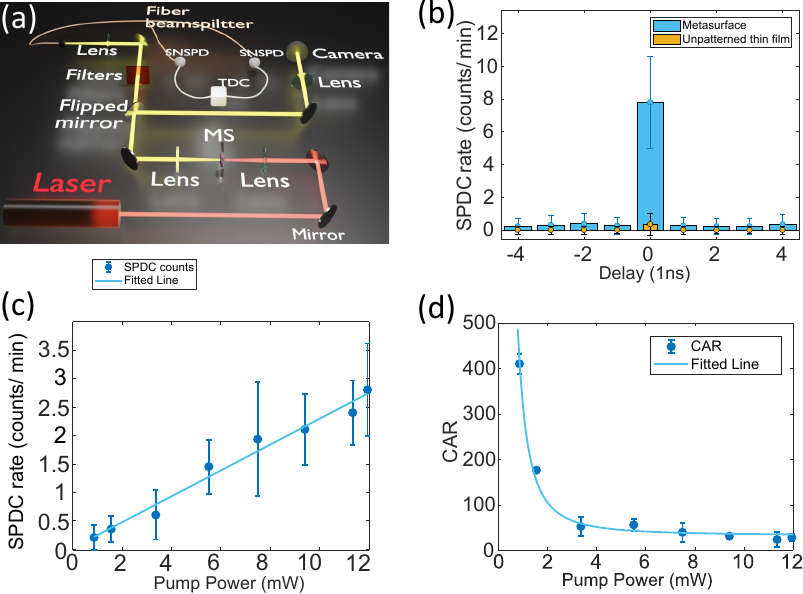} 
    \caption{
    \textbf{Confirmation of quantum photon pair generation from SPDC.}
    \textbf{a},~Experimental setup for SPDC-related measurements.
    \textbf{b},~Coincidence histograms for the metasurface and unstructured thin film with a pump of 12.5~mW at 770~nm, showing a 20-time enhancement from our metasurface compared to that from the unstructured thin film.
    \textbf{c},~The measured SPDC rate as a function of pump power and its linear fitting.
    \textbf{d},~Measured coincidence to accidental ratio (CAR) versus pump power.
}
\label{fig:3}
\end{figure*}

We further investigate photon-pair generation from the 3R-MoS$_2$ metasurface. 
All SPDC-related measurements were conducted using the setup shown in Fig.~\ref{fig:3}(a), with details provided in the methods section in supplementary (S8). 
The metasurface was pumped with a beam polarized along the y-direction. 
The SPDC rate is determined by the coincidence counts due to the simultaneous detection of two photons within a defined short time window.
The photon pairs are detected within a spectral range of 1500~nm to 1600~nm, corresponding to a bandwidth of 100~nm.
At resonance, the SPDC rate from the metasurface reached up to approximately 8 coincidence counts per minute when pumped at $2\times\lambda_{pump}=1540$~nm with 12~mW, as shown in the histogram in Fig.~\ref{fig:3}(b). 
The photon-pair generation rate from the metasurface was enhanced by approximately a factor of 20 compared to an unpatterned thin film under the same pump conditions (see Sec.~S5 in supplementary for detailed raw data). 
%The integration times for both the metasurface and the unpatterned film were set to 25 minutes.
At this pump power, the coincidence-to-accidental ratio (CAR) was around 32, indicating that the detected correlations indeed stem from SPDC. 
The CAR from the metasurface was higher than that from an unpatterned 3R-MoS$_2$ film~\cite{Weissflog:2024-7600:NCOM}, whose CAR is only around 16 based on our measurement. 
This is further reinforced by the measured dependence of the photon-pair rate on the pump power as shown in Fig.~\ref{fig:3}(c). 
All measurements were performed using the same integration duration.
We clearly found that the SPDC rate is linear with the pump power at $2\times\lambda_{pump}=1540$~nm, as expected for the spontaneous conversion process. 
We note a difference between the SPDC rate presented in Fig.~\ref{fig:3}(b) and Fig.~\ref{fig:3}(c) at the pump power of 12~mW.
This happens because the data in Fig.~\ref{fig:3}(b) were obtained using an optimized setup with higher detection efficiency.
%This linear relationship is crucial as it demonstrates the scalability of SPDC output with pump intensity, ensuring that the photon-pair generation rate can be enhanced with higher pump power~\cite{Zograf:2024-751:NPHOT}. 
%The 3R-MoS$_2$ phase exhibits an exceptionally high damage threshold, exceeding 45~GW/cm² under 1550~nm femtosecond pulsed laser irradiation.
%The structural and thermal properties of 3R-MoS$_2$ provide additional advantages for SPDC generation.  The atomic thinness of 3R-MoS$_2$ further aids in heat dissipation, reducing thermal effects that can limit SPDC efficiency in conventional materials~\cite{PhysRevB.94.115205}. 
The corresponding CAR, shown in Fig.~\ref{fig:3}(d), exhibits an inverse relationship with $1/P^2$, where $P$ is the power of the input pump beam. 
At the pump power of 0.85~mW, a CAR value of over 400 is observed, highlighting the high quality of the photon-pair source from the designed metasurface.

\begin{figure*}[!b] \centering
    \includegraphics[width=0.8\textwidth]{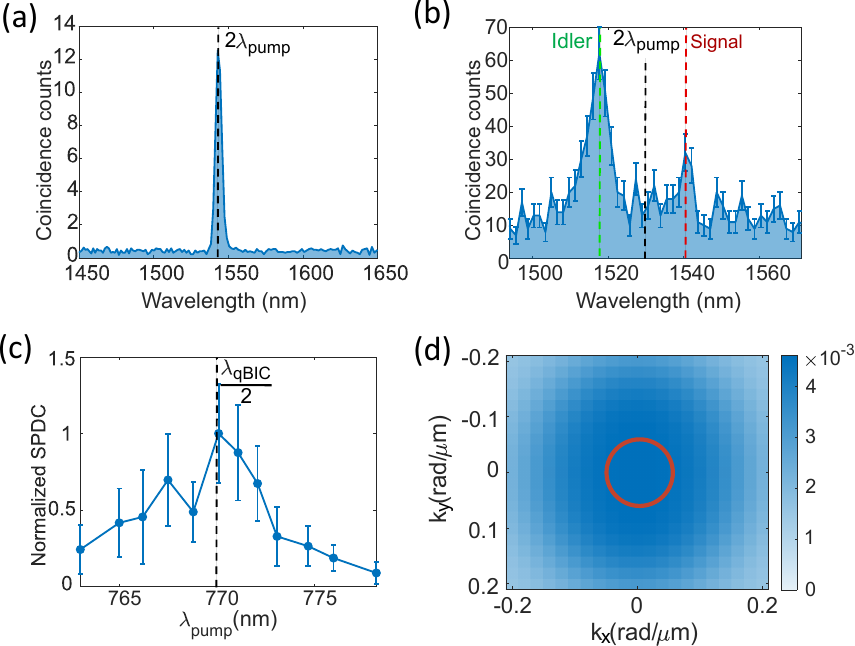} 
    \caption{ 
    \textbf{Quasi-BIC resonance enhanced quantum photon-pair generation.}
    \textbf{a},~Measured spectrum of the generated photon pairs with a 2-km long fiber. The bandwidth of the biphoton spectrum is narrow, which verifies the high quality factor of the resonant mode.
    The black dashed line indicates the position of $2\lambda_{pump}$, which overlaps with the qBIC resonance.
    \textbf{b},~Measured spectrum of the signal photons for a different pump frequency. 
    Nondegenerate photons can be generated due to the qBIC resonance.
    The black dashed line indicates the position of $2\lambda_{pump}$.
    Only the signal is emitted at the qBIC resonance (red dashed line).
    The spectrum agrees with the simulated nondegenerate SPDC (see Sec.~S3 in Supplementary).
    \textbf{c},~Measured SPDC rate versus pump wavelength.
    The dashed line indicates the position of $\lambda_{qBIC}/2$.
    \textbf{d},~Simulated emission pattern with a pump of 770.5~nm (unit: Hz$\cdot \mu$m$^2$/W). The SPDC rate is flat in the range up to 0.2~rad/$\mu$m ($\approx$3 degrees). The red circle indicates the area being detected in the experiment. The detection range is up to~0.05 rad/$\mu$m ($\approx$1 degree).
}
\label{fig:4}
\end{figure*}

To demonstrate the influence of the resonance on the spectral properties of the generated photon pairs, we measured the biphoton spectrum from our qBIC metasurface as shown in Fig.~\ref{fig:4}(a). 
To measure the spectrum, a long dispersive single-mode fiber of 2~km was inserted into the setup shown in Fig.~\ref{fig:3}(a)~\cite{Zielnicki2018, Valencia:2002-183601:PRL}. 
The pump laser used was again set at a wavelength of $2\times\lambda_{pump}=1540$~nm with 12~mW pump power. 
As shown in Fig. \ref{fig:4}(a), the spectrum is narrow with a linewidth
of around 4~
nm based on Fano fitting. 
The bandwidth is narrower than the linewidth observed in the linear transmission shown in Fig. \ref{fig:1}(e) because the SPDC rate is dependent on the product of the linear electric field in the nanoresonators.
This restriction of the spectral bandwidth stems from the resonant enhancement of the optical density of states, whereas for the unstructured MoS$_2$ crystal, a broad spectrum is expected. 
Thanks to the narrow bandwidth from our metasurface, a brightness enhancement greater than a factor of 20 at the resonant wavelength can be achieved. 
With a 100-nm detection bandwidth, the brightness enhancement is estimated to be approximately 400.
Additionally, the narrow bandwidth enables non-degenerate SPDC to generate more complex quantum states~\cite{Santiago-Cruz:2022-991:SCI}.

In Fig.~\ref{fig:4}(b), we show the measured emission spectrum of photon pairs for a different pump wavelength. 
By shifting the pump wavelength away from half of the qBIC resonance wavelength, one photon will be generated from the qBIC resonance, while the other photon is non-resonant.
In the experiment, the pump laser is tuned to at $\lambda_{pump}$=765~nm.
As indicated in the measured spectrum (Fig. \ref{fig:4}(b)), the signal photon has a wavelength around 1540 nm due to the qBIC resonance.
To satisfy the conservation of energy $1/\lambda_{signal}+1/\lambda_{idler} = 1/\lambda_{pump}$, the idler is around 1519~nm. 
This could be used for generating spectral entanglement, which is advantageous for compact quantum information processing techniques~\cite{Ladd2010Nature}.
Due to the inherent symmetry of the SPDC process, signal and idler photons are generated in pairs with equal probability for a given pump wavelength. Therefore, regardless of which photon is at $\lambda_{qBIC}$, the signal and idler photons should exhibit equal generation rates.
The discrepancy observed in the experimental coincidence spectrum, where the idler (off-resonant) peak appears higher than the signal (on-resonant) peak, is due to different detection efficiencies of the two detectors between the two spectral regions.
Similar height differences have been observed in other metasurface platforms~\cite{Santiago-Cruz:2022-991:SCI, Noh:2024-15356:NANL}.
%For the degenerate pump wavelength, we find one emission maximum as in the measured spectrum in Fig.~\ref{fig:4}(a).
%As we shift the pump wavelength $2\lambda_{pump}$ off the qBIC resonance, two narrow peaks appear in the biphoton spectrum. 
%One centered at the qBIC peak resonace ($\lambda_s = 1540$nm), while the other one varies depending on the pump wavelength to satisfy the conservation of energy: $\lambda_{\mathrm{i}}=\left(\frac{1}{\lambda_{\mathrm{p}}}-\frac{1}{\lambda_{\mathrm{s}}}\right)^{-1}$.
%One of the nondegenerate SPDC cases is observed experimentally at a pump wavelength of 765~nm (see measured spectrum in Supplementary).

The role of the resonance for enhancing SPDC emission is further reinforced by the dependence of the photon-pair rate on the pump power, shown in Fig.~\ref{fig:4}(c). At a pump wavelength of 770.5 nm, where energy conservation allows both signal and idler photons to be emitted at the resonantly enhanced wavelength, we find an enhancement of up to 10 compared to pump wavelengths that do not enable degenerate SPDC at the qBIC wavelength. This increase is again attributed to the qBIC resonance, which greatly enhances the nonlinear light-matter interaction within 3R-MoS$_2$.
%As a result, our designed 3R-MoS$_2$ our designed metasurface enables higher SPDC rates, making it well-suited for applications that demand high photon-pair generation rates, such as scalable quantum communication and integrated photonic devices.

The emission pattern from SPDC is simulated in COMSOL using the quantum-classical correspondence between the quantum SPDC process and the classical sum-frequency generation (SFG) process~\cite{Parry:2021-55001:ADP}. 
SFG can be regarded as the inverse process of SPDC, where two photons of lower frequencies combine to generate one photon of higher frequency. 
The resulting pattern with a vertically polarized pump at resonance ($2\times\lambda_{pump}=1540$~nm) is shown in Fig.~\ref{fig:4}(d). 
The generated signal and idler photons are both in the same polarization state ($\ket{VV}$) as the pump photon. 
Additionally, according to the emission pattern of $\ket{VV}$, the SPDC rate remains high in the range from -0.2~rad/$\mu$m to 0.2~rad/$\mu$m. 
The optimal SPDC rate measured in our experiment is approximately 0.01 Hz/mW. With a broader detection angle, this rate can be enhanced by an order of magnitude, reaching values comparable to state-of-the-art metasurfaces, such as the $[110]$-oriented InGaP metasurface reported in Ref.~\cite{Ma:2025-adu4133:SCA}.
This flat emission angle aligns with our linear simulation of angular dispersion shown in Fig.~\ref{fig:1}(c). 
For comparison, achieving a truly broad-angle photon-pair source in conventional material systems like lithium niobate (LN) remains challenging. 
The state-of-the-art in flat angular generation based is approximately 3 degrees, which is theoretically proposed through mode hybridization but not yet achieved experimentally~\cite{Yuxin:2025-26115:APLQ}. 
Therefore, such a flat SPDC emission pattern shows the advantage of VdW materials and benefits many quantum applications, such as quantum imaging~\cite{Ma:2025-2:ELI}. 
However, in the experiment, the detection range is limited to around 0.05 rad/$\mu$m (represented by a red circle in the figure), indicating that the rate can be further enhanced with a larger detection range. 
In contrast, the generation rate of other quantum states, where the signal and idler are in different polarization states ($\ket{HH}$, $\ket{HV}$, and $\ket{VH}$), is several orders of magnitude lower than the rate of $\ket{VV}$ (see Sec.~S1.5 in Supplementary). 
This reveals that the metasurface can tailor the state of generated photon pairs beyond the structure of the nonlinear tensor.

%\section*{Discussion}

This work presents, for the first time, a quantum light source from a metasurface composed of a van der Waals (VdW) crystal, 3R-MoS$_2$. 
The unique properties of 3R-MoS$_2$ offer several advantages for photon pair generation and related applications. 
Firstly, 3R-MoS$_2$ has a higher refractive index compared to other material systems~\cite{Zograf:2024-751:NPHOT, Sultanov:2024-294:NAT}, enhancing light trapping and broadening the generated angle range of photon pairs. 
Additionally, its atomically thin layer facilitates integration with compact quantum and optical devices~\cite{Sun2024CR}. 
There have also been demonstrations of more efficient SPDC from quasi-phase-matched 3R-MoS$_2$, which can be integrated into the metasurface to further enhance performance~\cite{Trovatello:2025-291:NPHOT, Tang2024NC}.
%Most importantly, the substantial second-order susceptibility from van der Waals materials, combined with geometric resonance, significantly boosts the photon pair generation rate.

The metasurface increases the photon pair generation rate by a factor of 20 compared to unpatterned thin films. 
The narrow bandwidth of the qBIC resonance enables ultrahigh brightness enhancement, estimated to reach 400 times. 
Integrating such a brightness-enhanced 3R-MoS$_2$ metasurface into existing optical and quantum devices could lead to more efficient and compact solutions for various technological applications, from telecommunications to imaging. 

These findings suggest that 3R-MoS$_2$ metasurfaces could play a crucial role in the development of next-generation quantum information technologies. 
Future research could explore optimizing the metasurface design for specific applications, such as quantum computing or secure communication systems. 
Additionally, investigating other VdW materials with similar properties could further enhance the versatility and performance of quantum light sources.
VdW materials, including 3R-MoS$_2$, can host tightly bound excitons that remain stable at room temperature. 
These excitonic resonances can significantly enhance the quantum photon-pair generation rate, in stark contrast to III-V semiconductors, where excitons typically only survive at cryogenic temperatures~\cite{Guo:2023-53:NAT}.
There is potential for further improvement, including the integration of an in-plane cavity and mode hybridization for broader flatband generation.

% Your references go at the end of the main text, and before the
% figures.  For this document we've used BibTeX, the .bib file
% scibib.bib, and the .bst file Science.bst.  The package scicite.sty
% was included to format the reference numbers according to *Science*
% style.

%BibTeX users: After compilation, comment out the following two lines and paste in
% the generated .bbl file. 

% \section*{Supplementary materials}

% Supplementary material for this article is available.\\
% Sections S1 to S3\\
% Figures S1 to S10\\
% Table S1\\% Coupled-mode theory coefficients.\\
% References
% %\textit{(45-54)}

%%%%%%%%%%%%%%%%%%%%%%%%%%%%%%%%%%%%%%%%%%%%%%%%%%%%%%%%%%%%%%%%%%%%%
%% The appropriate \bibliography command should be placed here.
%% Notice that the class file automatically sets \bibliographystyle
%% and also names the section correctly.
%%%%%%%%%%%%%%%%%%%%%%%%%%%%%%%%%%%%%%%%%%%%%%%%%%%%%%%%%%%%%%%%%%%%%
\bibliography{meta}

\section*{Supporting Information}
The Supporting Information is available free of charge at \url{https://pubs.acs.org/doi/full/10.1021/acs.nanolett.5c02170}.
The Supporting Information is available. It includes: Detailed theoretical derivations of tailored quantum states in metasurfaces (Section S1), Fano resonance analysis of linear transmission spectra (Section S2), methodology for generating nondegenerate photon pairs (Section S3), characterization of birefringence effects in sapphire substrates (Section S4), raw coincidence measurement data (Section S5), material properties of 3R-MoS\textsubscript{2} including second-order susceptibility and photoluminescence (Section S6), metasurface fabrication procedures (Section S7), and experimental setups for linear transmission, SHG, and SPDC (Section S8).

\section*{Data availability} 
All other data needed to evaluate the conclusions in the paper are present in the main text or the supplementary materials.

\section*{Acknowledgments}
This work was supported by the Australian Research Council grants DP190101559 (T.F., J.M., A.A.S.), CE200100010 (T.F., J.M., A.A.S.), DP240101011 (Y.L.), DP220102219 (Y.L.), DP180103238 (Y.L.), LE200100032 (Y.L.), ARC Centre of Excellence in Quantum Computation and Communication Technology project number CE170100012, and the Deutsche Forschungsgemeinschaft (DFG, German Research Foundation), projects Meta Active IRTG 2675 (437527638), NOA SFB 1375 (398816777), and MEGAPHONE (505897284). Sina Saravi acknowledges funding by the Nexus program of the Carl-Zeiss-Stiftung (project MetaNN). Metasurface fabrication was performed by Prof. Yuerui's group.

\section*{Author contributions}
All authors contributed to the experiment, the analysis of the results, and the writing of the manuscript.

\section*{Competing interests}
The authors declare no competing interests.

\end{document}